\documentstyle[aps,psfig,preprint,amssymb]{revtex}
\tightenlines
\begin{document}
\title{Adaptive Quantum Homodyne Tomography}
\author{Giacomo M. D'Ariano and Matteo G. A. Paris}
\address{{\sc Theoretical Quantum Optics Group}
\\Dipartimento di Fisica 
'Alessandro Volta' dell'Universit\`a di Pavia  \\  
Istituto Nazionale di Fisica della Materia -- Unit\`a di Pavia \\
{via Bassi 6 -- I-27100 Pavia, ITALY}}
\date{\today}
\maketitle
\begin{abstract}
An adaptive optimization technique to improve precision of 
quantum homodyne tomography is presented. The method is based on 
the existence of so-called {\em null functions}, which have zero 
average for arbitrary state of radiation. Addition of null functions 
to the tomographic kernels does not affect their mean values, but changes 
statistical errors, which can then be reduced by an optimization
method that "adapts" kernels to homodyne data.
Applications to tomography of the density matrix and other relevant 
field-observables are studied in detail. 
\end{abstract}
\section{Introduction}\label{s:intro}
The possibility of measuring the quantum state of radiation has been received
an increasing interest in the last years \cite{jmo,bil,pro}, as it opens
perspectives for a new kind of experiments in quantum optics, with the
possibility of measuring photon correlations on a sub-picosecond time-scale
\cite{mca}, characterizing squeezing properties \cite{sch}, photon statistics 
in parametric fluorescence \cite{cbs}, quantum
correlations in down-conversion \cite{kum}, nonclassicality of states
\cite{ncl}, and measuring Hamiltonians of nonlinear optical devices
\cite{lor}.  Among the many state reconstruction techniques suggested in the
literature \cite{ris,ray,tom,ulf,wod,buz,par,vog,lx0,opa}, quantum homodyne 
tomography (QHT) \cite{ray,tom,ulf,lx0} of radiation field have been
received much attention \cite{jmo}, being the only method which has been 
implemented in quantum optical experiments \cite{mca,sch,ray}, and recently 
being extended to estimation of the expectation value of any operator of the 
field \cite{lx0}, which makes the method the first universal detectors for 
radiation. 
\par
On one hand, QHT takes advantage of amplification from the local
oscillator in the homodyne detector, avoiding the need of single-photon 
resolving photodetectors, hence with the possibility of achieving very 
high quantum efficiency using
photodiodes \cite{kum}. On the other hand, the method of QHT is very efficient
and statistically reliable, and can be implemented on-line with the
experiment.
\par
In principle, a precise knowledge of the density matrix would require an
infinite number of measurements on identical preparations of radiation.
However, in real experiments one has at disposal only a finite number of data,
and thus statistical analysis and errors estimation are needed.  The purpose
of this paper is to analyze the possibility of improving the current QHT
technique, in order to minimize statistical errors. We will present a new
method that "adapts" the tomographic estimators to a given finite set of data,
improving the precision of the tomographic measurement.
\par
Quantum tomography of a single-mode radiation field consists of  a set of
repeated measurements of the field-quadrature $\hat x_{\phi}=\frac{1}{2}
(ae^{-i\phi}+a^{\dag}e^{i\phi})$ at different values of the reference phase 
$\phi$. The expectation value of a generic operator can be obtained by 
averaging a suitable kernel function $R[\hat O](x,\phi)$ as follows \cite{lx0}
\begin{eqnarray}
\langle\hat O\rangle\doteq\hbox{Tr}\left\{\hat\varrho\:\hat O\right\}=
\int_0^{\pi}\!\!\frac{d\phi}{\pi} \int_{-\infty}^{\infty} \!\! 
dx\;p_\eta(x,\phi)\;R_\eta[\hat O](x,\phi)\label{avprob}\;,
\end{eqnarray} 
where $p_\eta(x,\phi)$ denotes the probability distribution of the outcomes
$x$ for the quadrature $\hat x_{\phi}$ with quantum efficiency $\eta$, and 
$R_\eta [\hat O](x,\phi)$ is given by 
\begin{eqnarray}
R_\eta[\hat O](x,\phi) = \frac{1}{4}\int_{0}^{\infty} \!\! dr 
\exp\left[\frac{1-\eta}{8\eta}r\right] \:
\hbox{Tr}\left\{\hat O \:
\cos \left[\sqrt{r} (x-\hat{x}_{\phi})\right] \right\}
\label{kerdef}\;.
\end{eqnarray}
In the following we will focus attention only on the case $\eta=1$, and we
will drop the subscript $\eta$ in the notation. As it will appear from the 
following, the method works equally well also for nonunit quantum 
efficiency, and a
detailed numerical analysis versus $\eta$ will be given elsewhere.  On the
basis of identity (\ref{avprob}), it follows that the ensemble average
$\langle\hat O\rangle$ can be experimentally obtained by averaging 
$R[\hat O](x,\phi)$ over the set of homodyne data, namely 
\begin{eqnarray}
\langle \hat O \rangle = \overline{R[\hat O]} = \frac{1}{N}\sum_{i=1}^N
\;R[\hat O](x_i,\phi_i)\label{avdata}\;,
\end{eqnarray} 
N being the total number of measurements of the sample.  The statistical error of
the tomographic measurement in Eq. (\ref{avdata}) can be easily evaluated
provided that the corresponding kernel function satisfies the hypothesis of
the central limit theorem, which assures that the partial average over a block
of data is Gaussian distributed around the global average over all data. In
this case, the error is evaluated by dividing the ensemble of data into
subensembles, and calculating the r.m.s. deviation of each subensemble mean
value with respect to the global average.   The estimated value of such a
confidence interval is given by
\begin{eqnarray}
\delta O = \frac{1}{\sqrt{N}}\left\{\overline{\Delta 
R^2[\hat O]}\right\}^{1/2} \label{confin}\;,
\end{eqnarray}
where $\overline{\Delta R^2[\hat O]}$ is the variance of the kernel over 
the tomographic probability
\begin{eqnarray}
\overline{\Delta R^2[\hat O]} = \int_0^{\pi}\!\!\frac{d\phi}{\pi} 
\int_{-\infty}^{\infty}\!\!\! dx\;p(x,\phi)\;R^2[\hat O](x,\phi) 
- \left\{\int_0^{\pi}\!\!\frac{d\phi}{\pi}
\int_{-\infty}^{\infty}\!\!\! dx\;p(x,\phi)\;R[\hat O](x,\phi)\right\}^2 
\label{varker}\;.
\end{eqnarray}
Following this scheme, the tomographic precision in determining matrix
elements of the density operator $\hat\varrho$ has been discussed in 
\cite{ulf,com,nic}, with asymptotic estimations in Ref. \cite{las}, whereas 
relevant observables $\hat O$ have been analyzed in \cite{add}, also in 
comparison with the corresponding ideal measurement. 
\par\noindent 
The crucial point of the method presented in this paper is that the 
tomographic kernel $R[\hat O](x,\phi)$ is not unique, since a large class of
{\em null functions} \cite{dak,obs} $F(x,\phi)$ exists that have zero 
tomographic average for arbitrary state, namely 
\begin{eqnarray}
\overline{F}=\int_0^{\pi}\!\!\frac{d\phi}{\pi} 
\int_{-\infty}^{\infty} \!\! dx\;p(x,\phi)\;F(x,\phi) \; \equiv \; 0
\label{nuldef}\;.
\end{eqnarray}
Therefore, addition of null functions to a generic kernel gives a new kernel
with the same tomographic average, hence equivalent for the estimation of the
same ensemble average $\langle \hat O \rangle$.  On the
other hand, adding null functions would modify the kernel variance, whence the
statistical error over data.  The adaptive tomography method thus consists in
optimizing kernel in the equivalence class, in order to minimize the
statistical errors.
\par
The paper is structured as follows. In Section \ref{s:adap} we introduce the
classes of null functions that will be used in the paper, and describe the
adaptive optimization method in detail.  In Section \ref{s:tomo} we apply the
adaptive method to the tomography of the density matrix in the photon number
representation.  In Section \ref{s:obs} we analyze the improvement of
precision in tomographic measurement of some relevant field-observables.
Section \ref{s:sys} briefly describes the effects of systematic errors on the
effectiveness of the method. Finally, Section \ref{s:out} closes the paper by
summarizing the main results.
\section{Adaptive Tomography}\label{s:adap}
The following functions have vanishing tomographic expectation 
(\ref{nuldef})
\begin{eqnarray}
G_n^+ (x, \phi) = e^{i (1+n) 2\phi}\;g_+ (xe^{i\phi}) \qquad 
G_n^- (x, \phi) = e^{ - i (1+n) 2\phi}\:g_- (xe^{-i\phi}) 
\label{gnulldef}\;.
\end{eqnarray}
In Eq. (\ref{gnulldef}) $n\geq 0$ and $g_\pm(z)$ are analytic functions of $z$.
The set ${\cal G}$ of null functions defined in Eqs. (\ref{gnulldef}) forms a
vector space over $\mathbb C$, and each class ${\cal G}^\pm =\left \{G_n^\pm 
\right\}$ separately is closed under multiplication
(without inverse).
\par
In order to prove vanishing expectation (\ref{nuldef}) for $G_n^\pm (x,\phi)$ 
we consider the Taylor expansion of functions $g_\pm (xe^{i\phi})$
\begin{eqnarray}
g_\pm (xe^{i\phi})= \sum_{k=0}^{\infty} \: c_k^\pm \: x^k \: e^{\pm i k \phi}
\label{gexp}\;,
\end{eqnarray}
which allows to write 
\begin{eqnarray}
\overline{G_{n}^\pm} = \int_0^{\pi}\!\!\frac{d\phi}{\pi}
\int_{-\infty}^{\infty}\!\!\! dx\;p(x,\phi)\;
e^{\pm i (1+n) 2\phi} g_\pm (xe^{i\phi})=
\sum_{k=0}^{\infty} c_k^\pm \int_0^{\pi}\!\!\frac{d\phi}{\pi}
\:e^{\pm \:i\:(k+2+2n)\:\phi}\:\langle \hat x_{\phi}^k 
\rangle\label{null1}\;,
\end{eqnarray}
where $\langle \cdot\cdot\cdot \rangle$ denotes the usual ensemble average.  
Using the Wilcox decomposition formula \cite{wil} one can write
\begin{eqnarray}
\langle \hat x_{\phi}^k \rangle = \frac{k!}{2^k} \sum_{p=0}^{[[k/2]]}
\sum_{s=0}^{k-2p} \frac{\langle a^{\dag s} a^{k-2p-s}\rangle}{2^p p! s! 
(k-2p-s)!} \:e^{i\:(2p+2s-k)\:\phi}
\label{null2}\;,
\end{eqnarray}
where $[[x]]$ denotes the integer part of $x$. Eq. (\ref{null2}) together 
with the identity
\begin{eqnarray}
\int_0^{\pi}\frac{d\phi}{\pi}\:e^{i\:q\:\phi}=\left\{\begin{array}{cl} 
0 & \hbox{q even}\\1 & \hbox{q = 0}\\ \frac{2i}{\pi q}& \hbox{q odd}
\end{array}\right.
\label{null3}\;,
\end{eqnarray}
prove that
\begin{eqnarray}
\int_0^{\pi}\!\!\frac{d\phi}{\pi}
\: e^{\pm i (k+2+ 2 n)\phi} \:\langle \hat x_{\phi}^k 
\rangle =0\;, \qquad n\geq 0 \:, k\geq 0  
\label{gzero}\;,
\end{eqnarray}
hence
\begin{eqnarray}
\int_0^{\pi}\!\frac{d\phi}{\pi}
\int_{-\infty}^{\infty}\!\!\! dx\;p(x,\phi)\;
G_n^\pm (x,\phi ) = 0\;,\qquad n \geq 0 
\label{Gzero}\;,
\end{eqnarray}
namely $G_n^\pm(x,\phi)$ are null functions for $n\geq 0$.
\par\noindent
In the following, we will focus attention on three particular sets of null 
functions. The type-I null functions are obtained from Eq. (\ref{gnulldef}) 
by choosing $n=0$ and $g(xe^{i\phi})\equiv x^k e^{ik\phi}$ for a given $k$, 
and will be denoted by $F_k^I(x,\phi)$, namely 
\begin{eqnarray}
F_k^I(x,\phi) = x^k e^{i(k+2)\phi} \qquad k=0,1,...
\label{Inull}\;.
\end{eqnarray}
The type-II null-functions correspond to the simple choice 
$g(xe^{i\phi})\equiv 1$, i. e.
\begin{eqnarray}
F_n^{II}(\phi) = e^{i(1 + n)2\phi} \qquad n=0,1,...
\label{IInull}\;.
\end{eqnarray}
Finally, the type-III null functions are a kind of intermediate choice 
between type I and type II classes, and are defined as follows
\begin{eqnarray}
F_l^{III}(x,\phi) = x^{k[l]}\:e^{i( k[l] + 2 + 2 n[l])\phi} \qquad l=0,1,...
\label{IIInull}\;,
\end{eqnarray}
where $k[l]$ and $n[l]$ are given in Table \ref{t:III}.
In the following we will use the notation $F_k(x,\phi)$, dropping the type
index I-III, when the identity under consideration holds for all three types.
\par\noindent
Let us consider a generic real kernel $R[\hat O](x,\phi)$. 
By adding $M$ null functions keeping the kernel as real, we have a 
new kernel $K[\hat O](x,\phi)$
\begin{eqnarray}
K[\hat O](x,\phi)= R[\hat O](x,\phi) + \sum_{k=0}^{M-1} \mu_k F_k(x,\phi) 
+ \sum_{k=0}^{M-1} \mu_k^* F_k^*(x,\phi) \label{knull1}\;,
\end{eqnarray}
where $F_k(x,\phi) \in {\cal G}^+$, $F_k^*(x,\phi) \in {\cal G}^-$, and  
$\mu_k$ are complex coefficients. By definition we have $\overline{K[\hat O]}
=\overline{R[\hat O]}$, whereas the variance of the new kernel $K[\hat O]
(x,\phi)$ is given by
\begin{eqnarray}
\overline{\Delta K^2[\hat O]}=\overline{\Delta R^2[\hat O]} + 2 \left\{
\sum_{k,l=0}^{M-1} \mu_k \mu_l^* \overline{F_k F_l^*} +\sum_{k=0}^{M-1} 
\mu_k \overline{R[\hat O] F_k} + \sum_{k=0}^{M-1} \mu_k^*\overline{R[\hat O] 
F_k^*} \right\}\label{newvar}\;.
\end{eqnarray}
In deriving the above formula we use the fact that both 
${\cal G}^+$ and ${\cal G}^-$ are closed under multiplication.   
\par
The variance of the modified kernel function in Eq. (\ref{newvar}) 
can be minimized with 
respect to the coefficients $\mu_k$, leading to the linear set of equations 
\begin{eqnarray}
\sum_l \mu_l \: \overline{F_k F_l^*} = - \overline{R[\hat O] F_k^*}
\label{optvar}\;.
\end{eqnarray}
It is convenient to rewrite the optimization equations (\ref{optvar}) in 
matrix form as follows
\begin{eqnarray}
{\bf A} \; {\bf \mu} = {\bf b} 
\label{optlin}\;.
\end{eqnarray}
where ${\bf A}$ is the Hermitian $M\times M$ matrix 
\begin{eqnarray}
A_{kl}=\overline{F_k F_l^*}
=\int_0^{\pi}\!\frac{d\phi}{\pi}
\int_{-\infty}^{\infty}\!\!\! dx\;p(x,\phi)\;
F_k(x,\phi) F_l^*(x,\phi)
\label{amat}\;,
\end{eqnarray}
and ${\bf b}$ is the complex vector
\begin{eqnarray}
b_k=- \overline{R[\hat O] F_k^*} 
= - \int_0^{\pi}\!\frac{d\phi}{\pi}
\int_{-\infty}^{\infty}\!\!\! dx\;p(x,\phi)\;
R[\hat O](x,\phi)\: F_k^*(x,\phi) 
\label{bvec}\;.
\end{eqnarray}
Notice that the vector ${\bf b}$ depends on both the kernel $R[\hat O]$ and 
the state $\hat\varrho$ under examination, whereas the matrix ${\bf A}$ 
depends on the state only. 
\par\noindent 
By substituting Eq. (\ref{optvar}) in Eq. (\ref{newvar}) and inverting 
Eq. (\ref{optlin}) we obtain 
\begin{eqnarray}
\Delta^2[\hat O] \doteq \overline{\Delta R^2[\hat O]}
-\overline{\Delta K^2[\hat O]} = 2\sum_{k,l=0}^{M-1} \mu_k \: 
A_{kl}\: \mu_l^* = 2\sum_{k,l=0}^{M-1} b_k \:\left(A^{-1}\right)_{kl} 
\:b_l^*  \geq 0 
\label{redvar}\;,
\end{eqnarray}
which expresses the variance decrease in terms of $\bf A$ and 
$\bf b$. \par
\par\noindent
Let us summarize the optimization procedure for the kernel $R[\hat O](x,
\phi )$. After collecting an ensemble of $N$ tomographic data, the 
quantities $\bf A$ and $\bf b$ are evaluated as tomographic experimental 
averages. Then, by solving the linear system (\ref{optlin}) one obtains 
the coefficients $\mu_k$ which are used to build the optimized kernel 
$K[\hat O](x,\phi )$. At this point, the same data set is used to average 
$K[\hat O](x,\phi )$ and, upon dividing the set into subensembles, the 
experimental error is evaluated, whose square now is reduced by the 
quantity $\Delta^2 [\hat O]/N$.
\par\noindent
The actual precision improvement of the tomographic measurement 
depends both on the state under examination (which affects both ${\bf b}$ 
and ${\bf A}$) and on the operator $\hat O$, whose kernel enters only in 
the expression of ${\bf b}$. An explicit expression for $A_{kl}$ can be 
obtained by means of Eq.(\ref{null2}), and generally depends on the type 
of null function that are involved. For type-II null functions 
it reduces to the identity matrix, independently on the state
\begin{eqnarray}
A_{kl}^{II}=\delta_{kl} \qquad \hbox{type-II null functions}\label{matII}\;,
\end{eqnarray}
$\delta_{kl}$ denoting Kronecker delta. For type-I null functions 
one has 
\begin{eqnarray}
A_{kl}^I = \frac{(k+l)!}{2^{k+l}} \sum_{p=0}^{min(k,l)} 
\frac{\langle a^{\dag l-p} \: a^{k-p}\rangle}{2^p\:p! (l-p)! (k-p) !}
\qquad \hbox{type-I null functions}
\label{matform}\;.
\end{eqnarray}
The explicit expression for coherent and Fock states is
\begin{eqnarray}
A_{kl}^I &=& \alpha^{k-l} \frac{(k+l)!}{k!}\:2^{-k-2l} L_l^{k-l} 
(-2|\alpha |^2)  \qquad\hbox{coherent state $|\alpha\rangle$} \quad 
(k\geq l) \\ 
A_{kl}^I &=& \delta_{kl} \: \frac{2^{k-n+1}}{n!\sqrt{\pi}}    
\int_0^\infty dy\: e^{-y^2}\: y^{2k} H^2_n (y)
\qquad\hbox{Fock state $| n \rangle$}
\label{matsome}\;,
\end{eqnarray}
where $H_n (x)$ denotes Hermite polynomials. 
Notice that for Fock states the matrix is diagonal 
(which is true also for type-II and type-III null 
functions).
\section{Adaptive tomography of the density matrix}\label{s:tomo}
In this section we apply the adaptive method to the tomographic measurement 
of the density matrix in the photon number representation. We evaluate the 
variance reduction $\Delta^2[|n\rangle\langle m|]$ in 
Eq. (\ref{redvar}) for $\hat O = |n\rangle\langle m|$ corresponding to
the tomographic measurement of the matrix elements $\varrho_{nm} = \langle 
m|\hat\varrho |n\rangle$. We consider the different types of null functions, 
and calculate $\Delta^2[|n\rangle\langle m|]$ versus the number $M$ of added 
null functions, for either coherent states, squeezed vacuum, Fock states, 
and the "Schr\"{o}dinger-cat" like superposition of coherent states given by
\begin{eqnarray}
|\psi\rangle = \frac{1}{2 \sqrt{ 1+\exp (-2|\alpha |^2)}}
\Bigg[ |\alpha\rangle + | - \alpha\rangle \Bigg]
\label{cat}\;.
\end{eqnarray}
In order to see the new adaptive method at work Monte Carlo simulated 
experiments are presented. 
\par\noindent 
Tomographic kernels for the matrix elements in the Fock basis have been 
firstly presented in Ref. \cite{tom}, with extension to non unit quantum
efficiency in Ref. \cite{ulf}, and factorization identities for the kernel 
in Ref. \cite{opt}. 
However, none of the above methods allows for an explicit analytical 
evaluation of the vector ${\bf b}$ in Eq. (\ref{bvec}). 
For this reason, we compute $\Delta^2[|n\rangle\langle m|]$ numerically
presenting results in terms of the relative variance reduction $\gamma$, 
defined as follows
\begin{eqnarray}
\gamma = 1 - \frac{\overline{\Delta K^2[\hat O]}}
{\overline{\Delta R^2[\hat O]}} = 
\frac{\Delta^2[\hat O]}{\overline{\Delta R^2[\hat O]}}
\label{gamma}\;.
\end{eqnarray}
A complete removal of fluctuations would correspond to $\gamma =1$. 
\subsection{Coherent States}
\begin{figure}
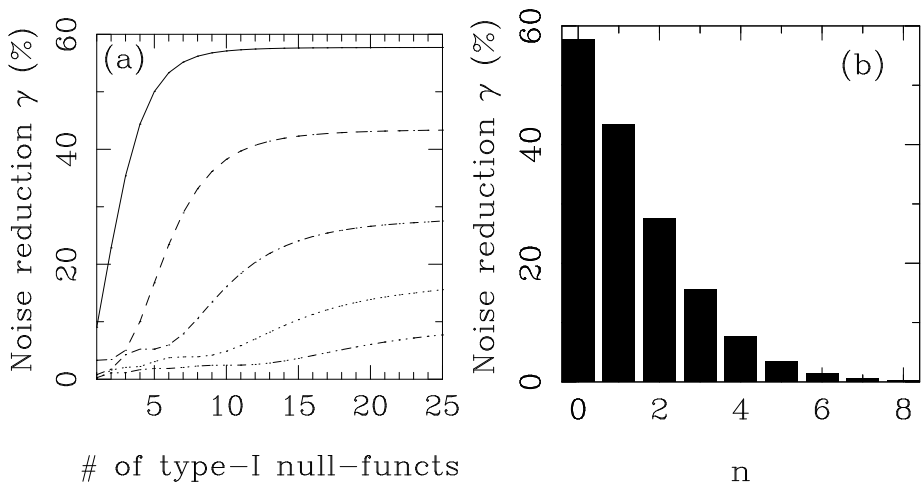
                                   
\begin{tabular}{cc}
\psfig{file=adap1a.ps,width=6cm} & \psfig{file=adap1b.ps,width=6cm}
\end{tabular}
\caption{Noise reduction in the tomographic measurement of the diagonal 
elements $\langle n|\hat\varrho |n\rangle$ of the density matrix of a coherent 
state $|\alpha\rangle$ with intensity $|\alpha |^2=5$. In (a): noise reduction 
$\gamma$ versus the number of added type-I null functions: the full curve 
represents $\langle 0|\varrho|0\rangle$, the dashed curve $\langle 1|\varrho
|1\rangle$, and so on, from the top to the bottom. In (b): noise 
reduction versus the index $n$ of the diagonal matrix element 
for $M=30$ added null functions.\label{f:cdiag1}}                     
\end{figure}

\begin{figure}
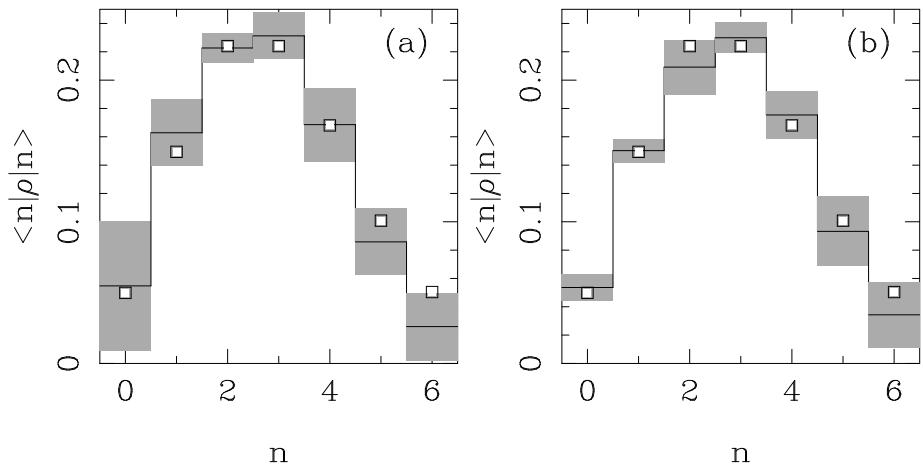
                                   
\begin{tabular}{cc}
\psfig{file=adap2a.ps,width=6cm} & \psfig{file=adap2b.ps,width=6cm}
\end{tabular}
\caption{Monte Carlo simulation of adaptive tomography of a coherent 
state with intensity $|\alpha |^2= 3$. A sample of $5$ blocks of 
$50$ homodyne data is used for each of $25$ phases (for a total 
number of measurements equal to $6250$). The optimization has been 
performed by adding $M=6$ null functions. In (a) the measured 
diagonal matrix elements before optimization, and in (b) after optimization. 
The squares indicate theoretical values. \label{f:csim}}                    
\end{figure}

The adaptive method leads to a significant error reduction for detection 
of matrix elements $\langle m | \hat\varrho |n\rangle$  of coherent states.
Our results indicates that type-I null functions are the most effective, and 
that the larger is the amplitude $\alpha$ of the coherent state, the larger 
the noise reduction.
In Fig. \ref{f:cdiag1} numerical results are presented
for diagonal elements $\langle n|\hat\varrho |n\rangle$ for intensity 
$|\alpha|^2 =5$. In Fig. \ref{f:cdiag1}(a) the noise reduction $\gamma$ is 
given versus the number $M$ of added type-I null functions. One can see
that the noise reduction $\gamma$ saturates for large $M$, and better levels 
$\gamma$ of reduction are achieved for smaller $n$.
In Fig. \ref{f:cdiag1}(b) the noise reduction is reported versus $n$ for 
$M=30$. In Fig. \ref{f:csim} we report the results from a Monte Carlo 
experiment for $|\alpha |^2= 3$, with optimization performed with $M=6$ 
null functions. The reduction of statistical errors for low values of $n$ 
is evident.
\par
\begin{figure}                                   
\begin{tabular}{ccc}
\psfig{file=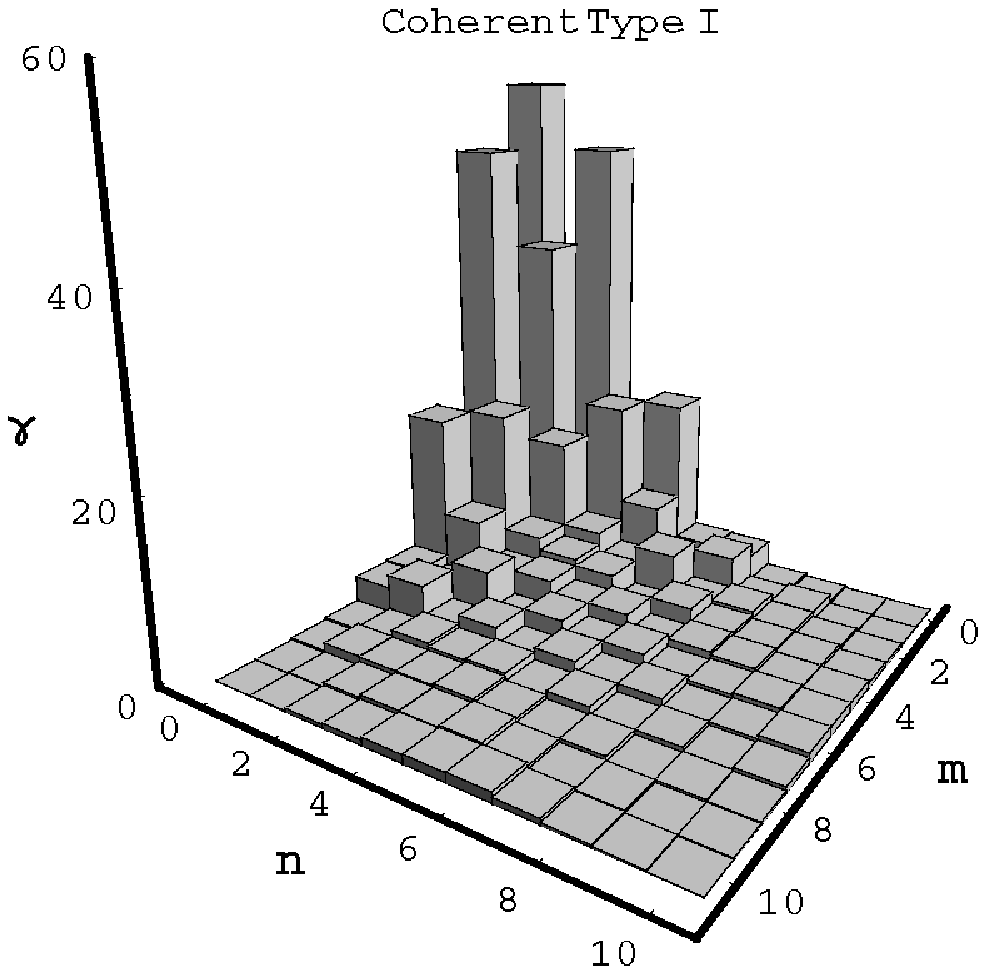,width=6cm}&
\psfig{file=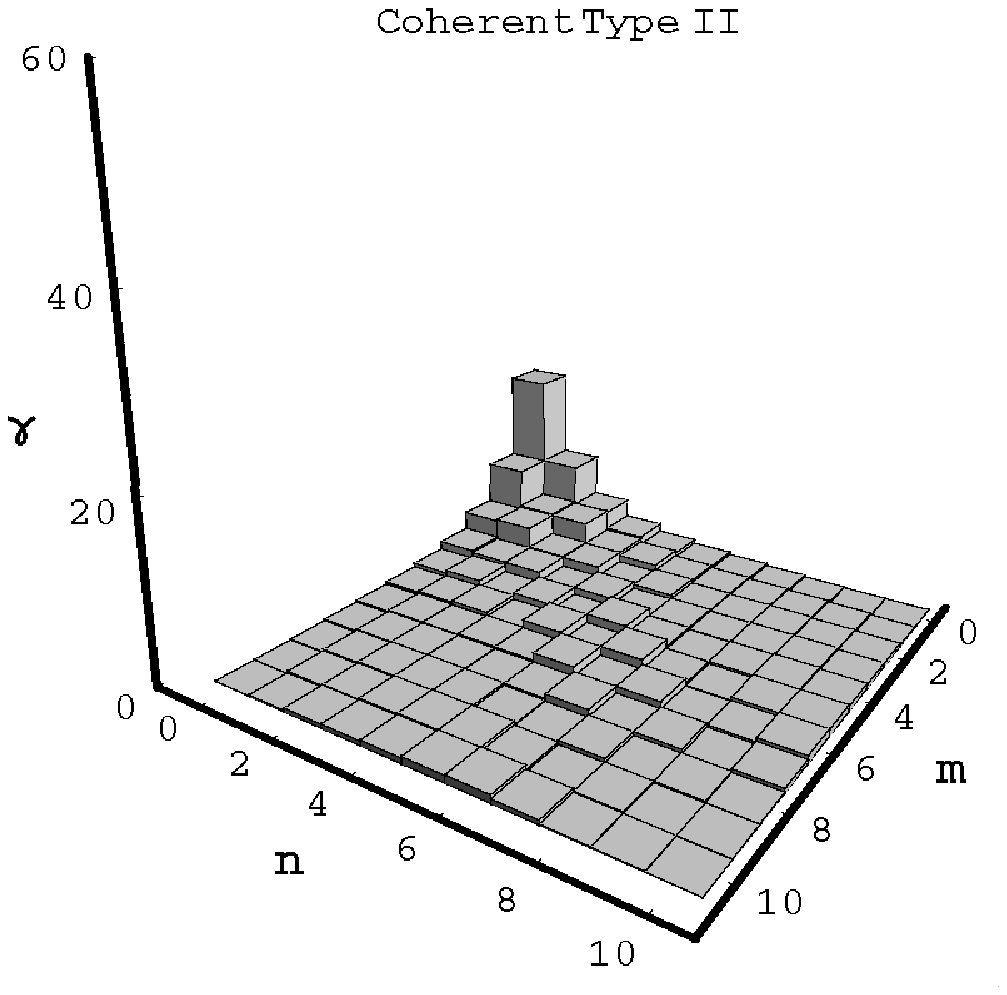,width=6cm}&
\psfig{file=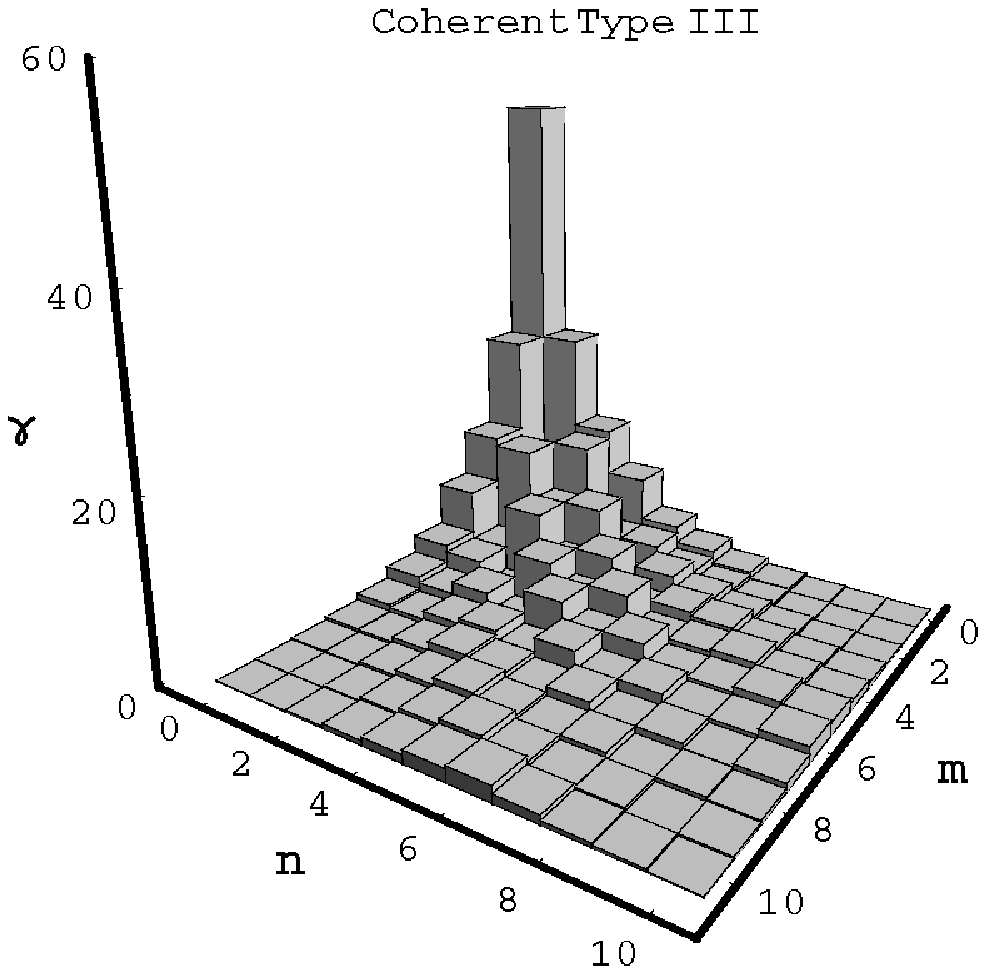,width=6cm}
\end{tabular}
\caption{Noise reduction $\gamma$ versus indices $n$ and 
$m$ of the matrix element $\langle m|\hat\varrho |n\rangle$ for a coherent 
state with intensity $|\alpha |^2=5$: (a) using only type-I null 
functions, (b) using only type-II, and (c) type-III.
For all plots $M=10$ null functions have been used 
in the optimization procedure.\label{f:coff}}                     
\end{figure}

\noindent
The noise reduction for the off-diagonal matrix elements behaves
similarly to the diagonal ones, being more effective for low indices. 
In Fig. \ref{f:coff} the noise reduction $\gamma$ versus $n$ and
$m$ of the matrix element $\langle m|\hat\varrho |n\rangle$ is plotted for
a coherent state with $|\alpha |^2=10$, and for the three 
types of null functions. The type-I null functions are generally 
more effective, though not uniformly over all indices $n$ and $m$. 
\subsection{Squeezed states and Schr\"{o}dinger cat states}
\begin{figure}                   
\begin{tabular}{ccc}
\psfig{file=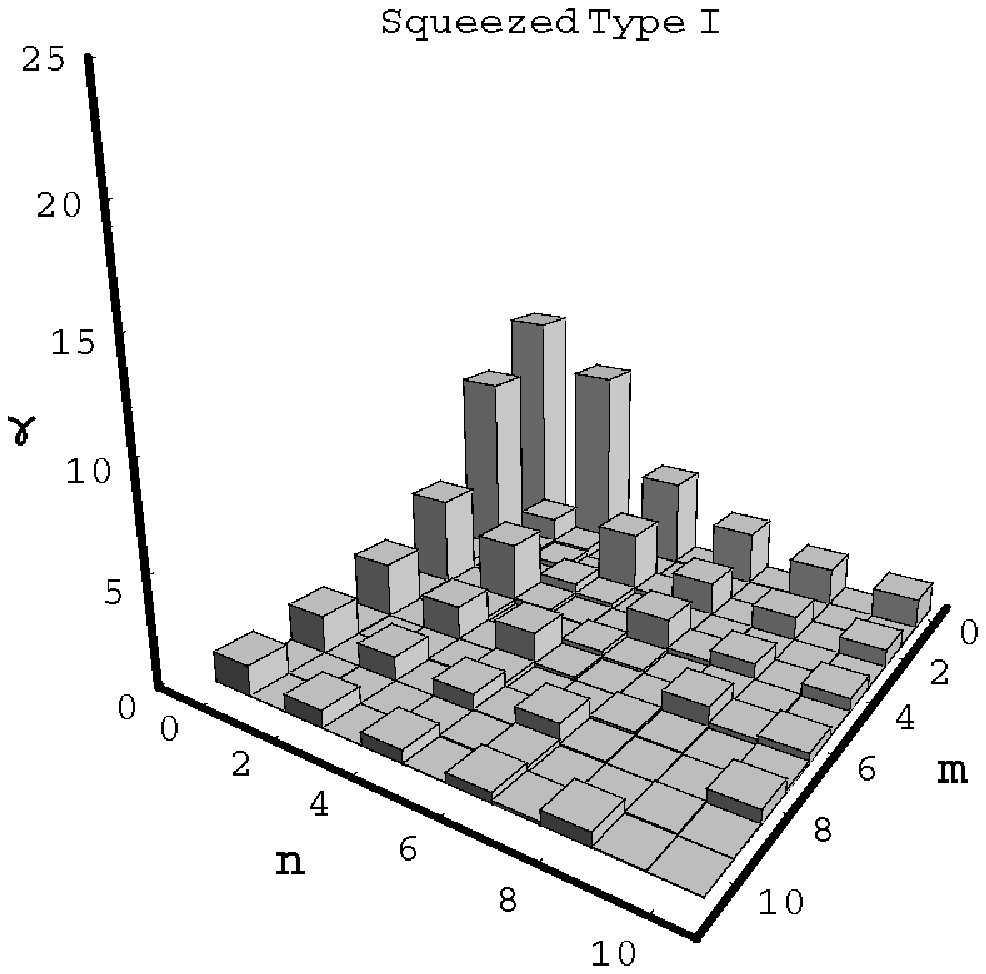,width=6cm}&
\psfig{file=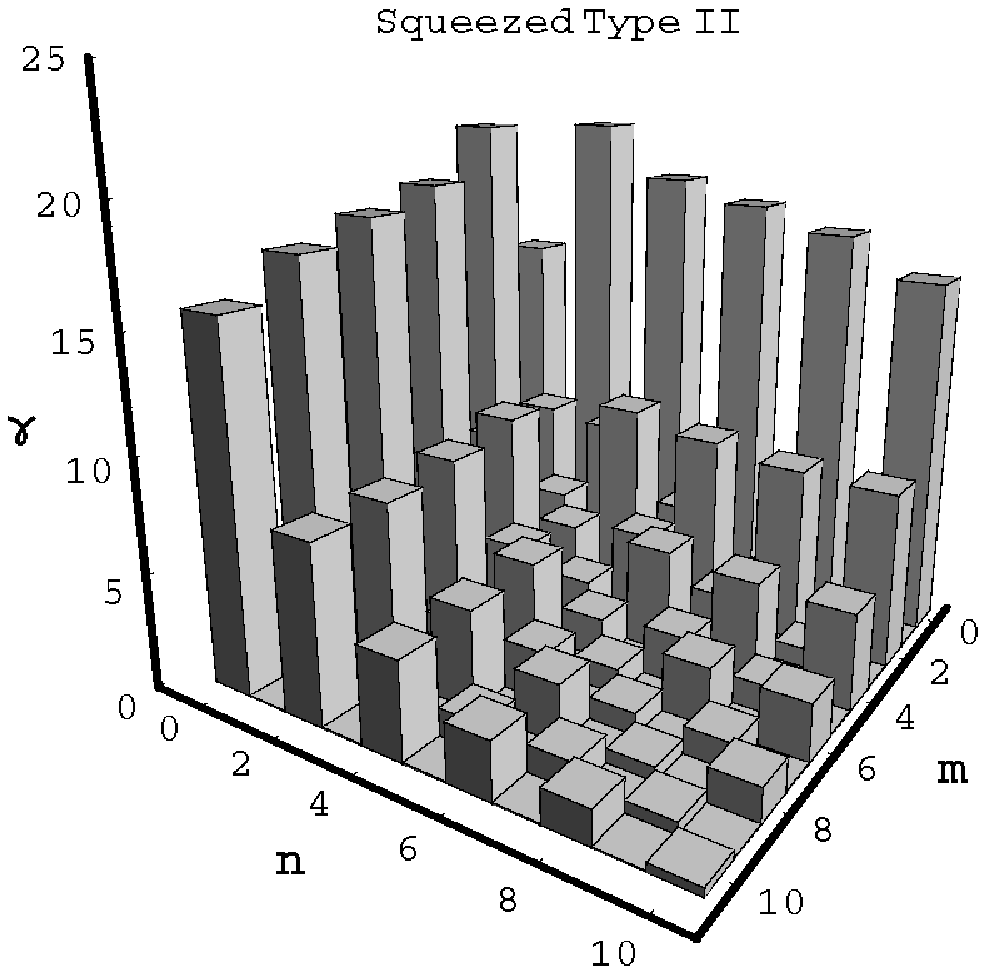,width=6cm}&
\psfig{file=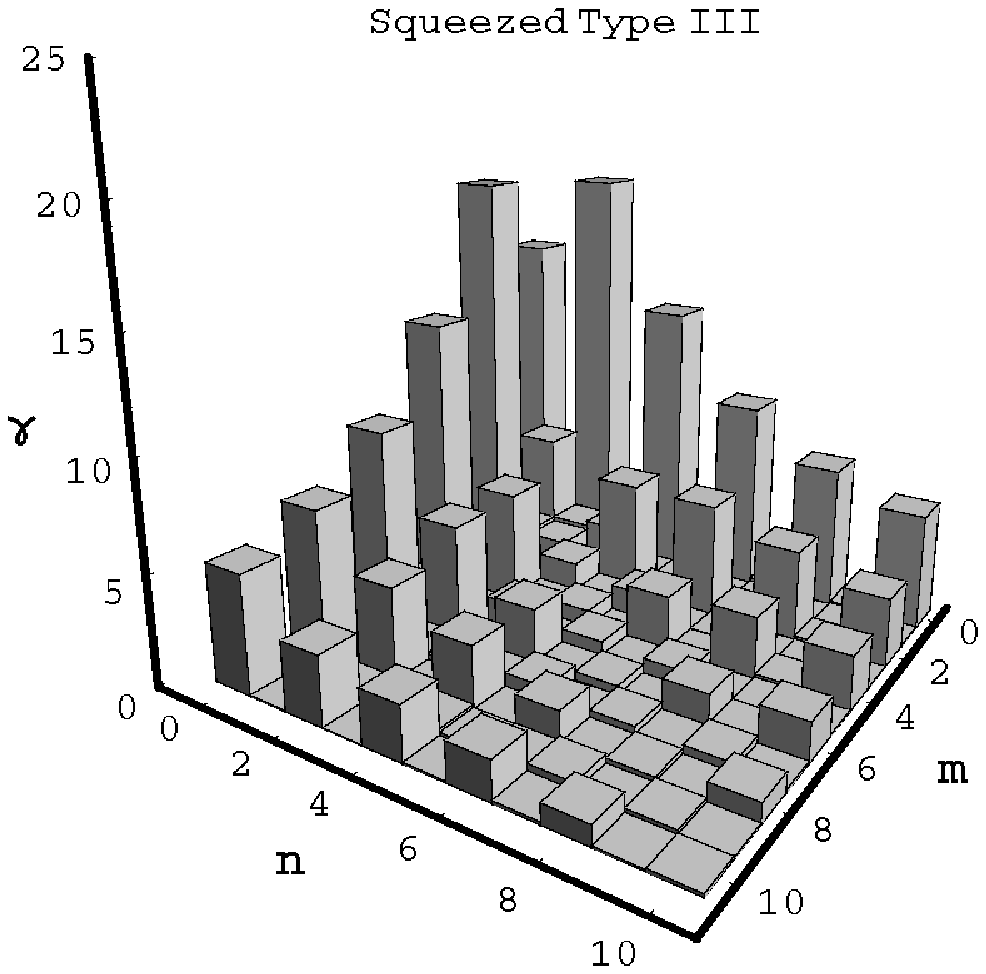,width=6cm}
\end{tabular}
\caption{Noise reduction for squeezed vacuum with $\langle\hat n \rangle =4$: 
(a) using only type-I null functions, (b) using only type-II, and (c) 
type-III. For all plots $M=10$ null functions have
been used in the optimization procedure.\label{f:catsw1}}
\end{figure}

\begin{figure}                   
\begin{tabular}{ccc}
\psfig{file=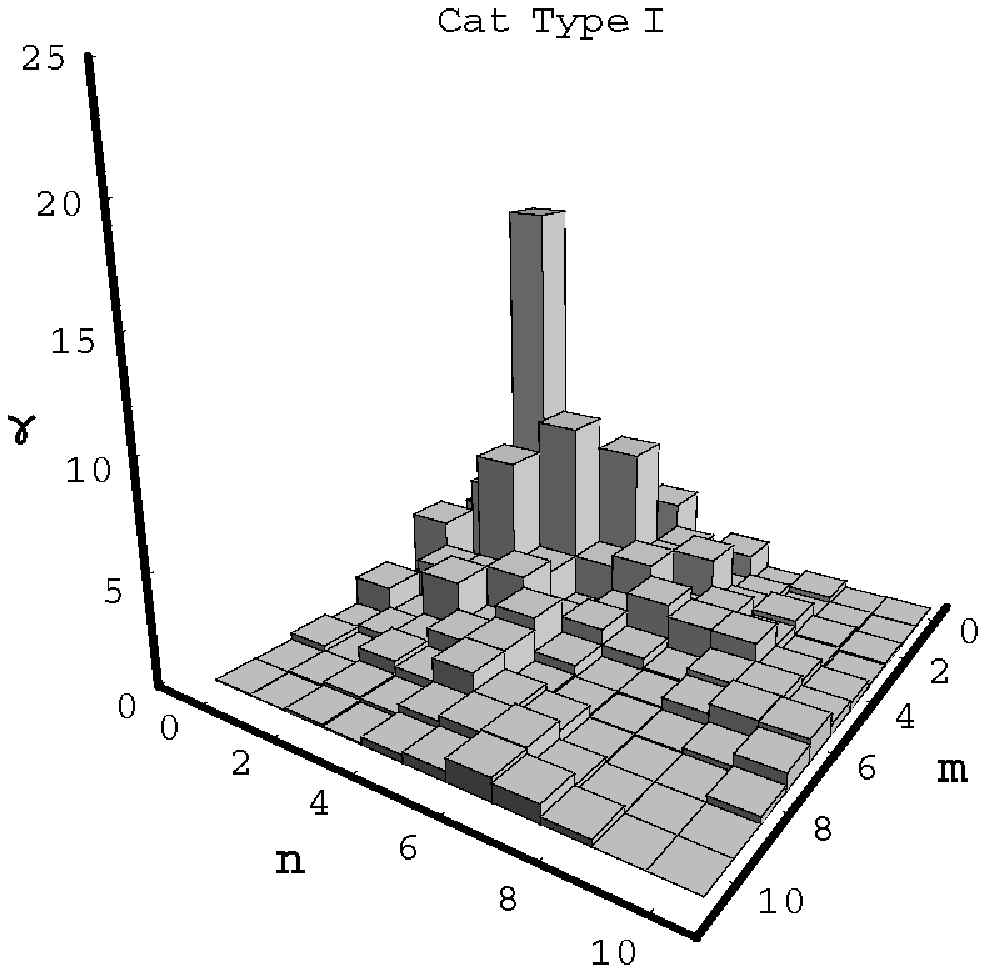,width=6cm}&
\psfig{file=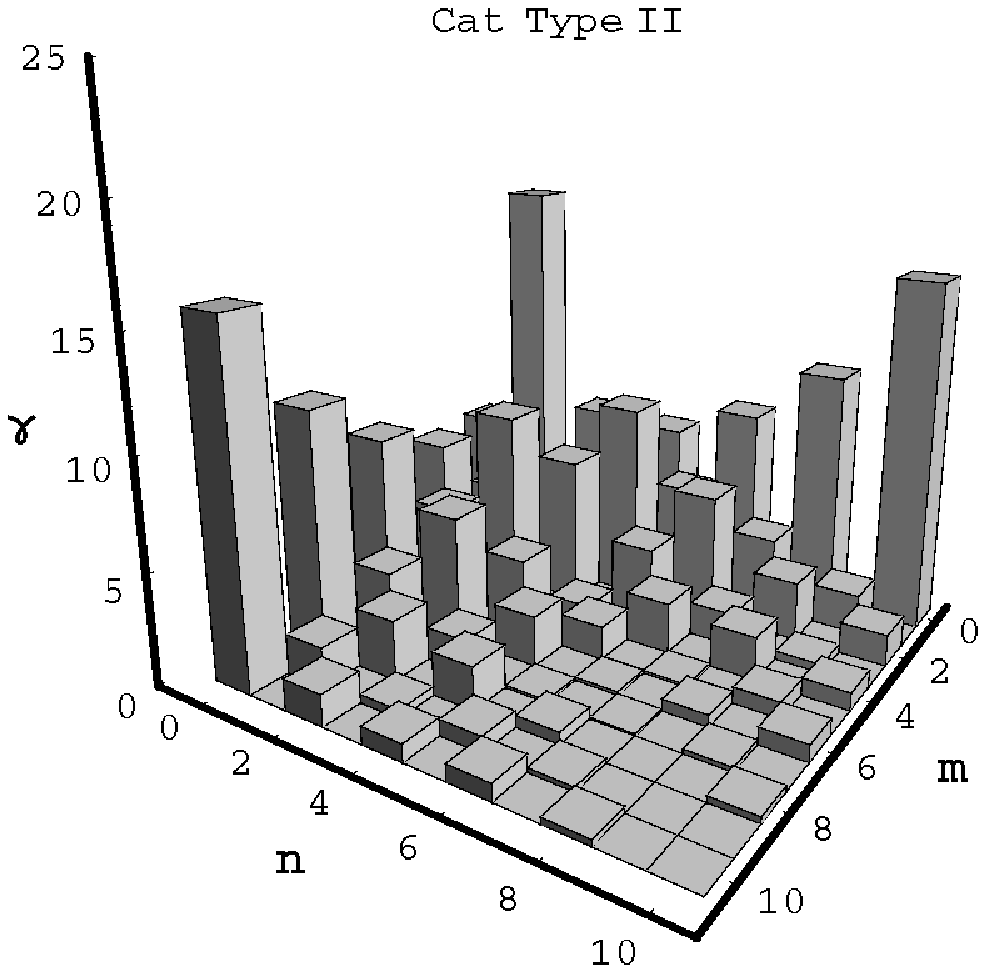,width=6cm}&
\psfig{file=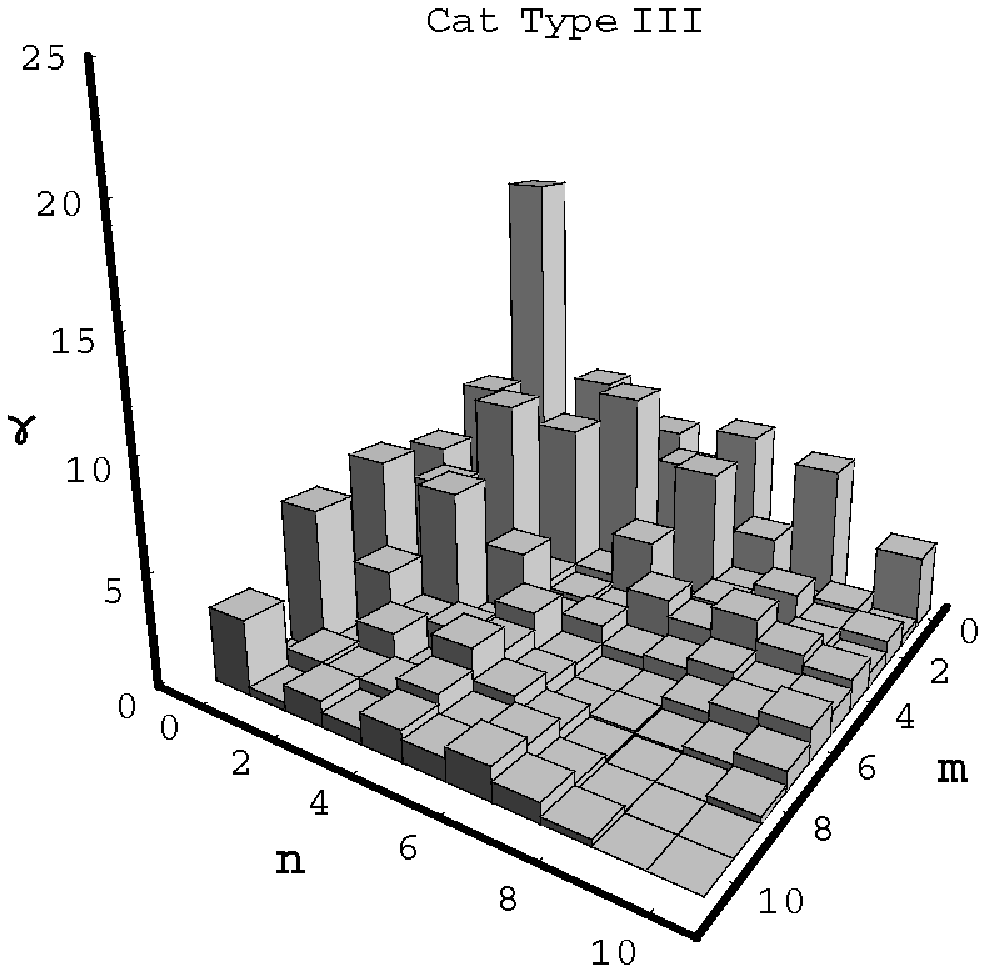,width=6cm}
\end{tabular}
\caption{Noise reduction for the cat-like superposition 
of coherent states and for the three types of null functions in Eq.
(\ref{cat}) with $\alpha=\sqrt{3}$: (a) using type-I null functions, 
(b) using type-II, and (c) type-III. For all plots $M=10$ null functions 
have been used in the optimization procedure. \label{f:catsw2}}
\end{figure}

\begin{figure}
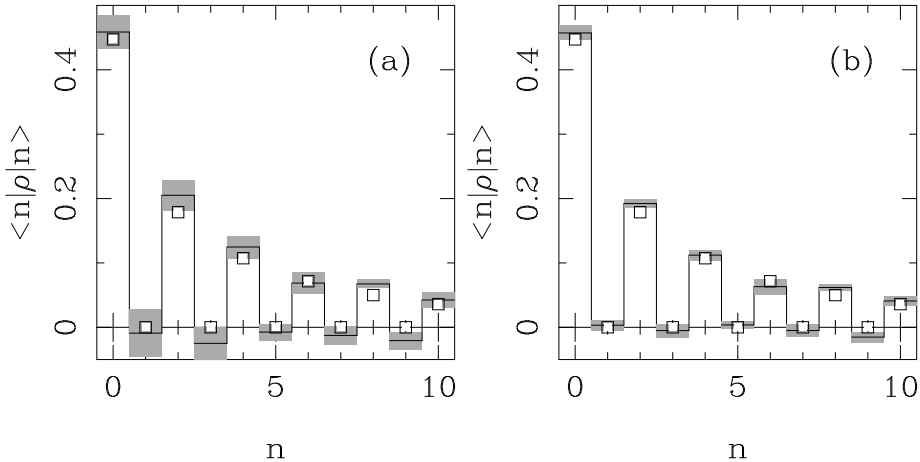

\begin{tabular}{cc}
\psfig{file=adap6a.ps,width=6cm} & \psfig{file=adap6b.ps,width=6cm}
\end{tabular}
\caption{Adaptive tomography of a squeezed vacuum with $\langle \hat n 
\rangle =4$. The Monte Carlo sample includes $5$ blocks of 
$100$ data for each of $50$ phases (for a total number of measurements 
equal to $25000$). The optimization has been performed by adding 
$M=10$ type-II null functions. (a) measured elements without 
optimization; (b) with optimization. The squares indicate the 
theoretical values.\label{f:sqsim}}
\end{figure}

Results for squeezed states and "cat" superposition of coherent states 
are presented in the same subsection, since they behave similarly. 
This is due to the fact that both states have phase-dependent 
features, that reflect in a similar odd-even oscillation in the photon 
number probability distribution. In Figs. \ref{f:catsw1} and \ref{f:catsw2} 
the noise reduction for both cases is plotted for the three types of 
null functions, for $M=10$. From the plots it is apparent that type-II 
null functions are now the most effective ones, especially for off-diagonal 
matrix elements, though the same level of noise reduction for low $n$ and 
$m$ can also be obtained using type-I and type-III null functions. 
In Fig. \ref{f:sqsim} results from a Monte Carlo simulated adaptive tomography
on a squeezed vacuum are reported for $\langle \hat n \rangle =4$ and $M=10$.
Matrix elements before and after optimization can be compared, showing
the error reduction at work.
\subsection{Fock states}
For Fock states the matrix ${\bf A}$ is diagonal for all types of null 
functions, and therefore the optimization procedure just consists of 
the evaluation of the vector ${\bf b}$. The kernels for the matrix 
elements have the form
$K[|n\rangle\langle m|] (x,\phi) 
=f_{n,m}(x)\exp(i(n-m)\phi)$, where $f_{n,m}(x)$
has the parity of $n-m$ \cite{bil,opt}. This fact, together with the 
integral  (\ref{null3}) makes straightforward to show that
\begin{eqnarray}
b^I_k \equiv b^{II}_k \equiv b^{III}_k \equiv 0 \qquad \forall k
\label{nofock}\;,
\end{eqnarray}
namely no improvement should be expected for the precision of quantum
tomography on Fock states.
\section{Adaptive tomographic measurements of observables}\label{s:obs}
The tomographic estimation of the ensemble average $\langle\hat O \rangle$
of a radiation operator $\hat O$ can be obtained by averaging the kernel
$R[\hat O](x,\phi)$ given in Eq. (\ref{kerdef}). However, Eq. (\ref{kerdef})
needs a procedure that exploits the null function equivalence, and is given
in Ref. \cite{lx1}. For this reason, for simplicity here we use the Richter 
formula \cite{ric}, which expresses the kernels for the normally ordered 
moments as follows
\begin{eqnarray}
R[a^{\dag}{}^n a^m](x;\phi)=e^{i(m-n)\phi}
\frac{H_{n+m}(\sqrt{2}\:x)}{\sqrt{2^{n+m}}{{n+m}\choose n}}\label{ric}\;,
\end{eqnarray}
$H_n(x)$ being the Hermite polynomial of order $n$. We apply the adaptive 
method to the tomographic detection of the most relevant observables: 
intensity, quadrature and complex field amplitude. The optimization method is here particularly useful, as the tomographic detection of these observables 
using the Richter kernel is very noisy \cite{add,rou}. 
\par\noindent
In contrast to the case of matrix elements given in Section \ref{s:tomo}, 
here some analytical evaluations can be carried out. We consider measurements 
performed on coherent states, squeezed vacuum, Fock states and cat 
superposition of coherent states.
It turns out that addition of just few null functions to the Richter kernels
generally results in a large improvement of the tomographic precision, 
again with the exception of Fock states where no improvement can be obtained. 
\subsection{Intensity}
The tomographic detection of intensity is obtained by averaging 
the kernel
\begin{eqnarray}
R[a^{\dag} a](x)=2 x^2 - \frac{1}{2}\label{num}\;.
\end{eqnarray}
The vectors ${\bf b}$ needed for the optimization procedure
are given by
\begin{eqnarray}
b_k^{I}= - \overline{R[a^{\dag} a] F_k^{I*}} = 
-\overline{2 x^{k+2} e^{-i(k+2)\phi}} &=& -\frac{\langle a^{\dag (k+2)}
\rangle}{2^{1+k}} \label{vecnumI} \\
b_k^{II}= - \overline{R[a^{\dag} a] F_n^{II*}} = 
-\overline{2 x^2 e^{-i(n+1)2\phi}} &=&
-\left\{\begin{array}{cl}
\frac{\langle a^{\dag 2}\rangle}{2} & n=0 \\
0 & n \neq 0 
\end{array}\right. \label{vecnumII} \\
b_k^{III}= - \overline{R[a^{\dag} a] F_l^{III*}} = 
-\overline{2 x^{k[l]+2}  e^{-i(k[l]+2+n[l])\phi}} &=&
-\left\{\begin{array}{cl}
\frac{\langle a^{\dag 2}\rangle}{2} & l=0 \\
0 & l \neq 0 
\end{array}\right. \label{vecnum}\;.
\end{eqnarray}
From Eqs. (\ref{vecnumII}) and (\ref{vecnum}) it follows that 
only $F_0^{I}(x,\phi)$ and $F_0^{II}(\phi)
\equiv F_0^{III}(\phi)\equiv \exp (2i\phi)$ are
effective in reducing the variance.  We solved 
analytically the optimization equations (\ref{optlin}) for type-I 
null functions, and also in this case it turns out that for all the 
states here considered, only the single null function $F_0^I(\phi)$ 
is needed, namely one has
\begin{eqnarray}
\mu_0 = b_0 \qquad \mu_k=0\:, \quad \forall\:k \geq 1
\label{optnum}\;.
\end{eqnarray}
The corresponding reduction of variance is easily obtained from 
Eq. (\ref{redvar}), and is given by 
\begin{eqnarray}
\Delta^2 [a^\dag a] =\frac{1}{2} \:\langle a^{\dag 2}\rangle\:
\langle a^2\rangle 
\label{rednum}\;.
\end{eqnarray}
Actually, $\Delta^2 [a^\dag a]$ can compensate the leading term of 
the variance of the original Richter kernel \cite{add}, which, in turn, is
given by
\begin{eqnarray}
\overline{\Delta R^2[a^\dag a]} =\langle\widehat{\Delta n^2}\rangle 
+ \frac{1}{2} \left[\langle a^{\dag 2}\:a^2\rangle + 
2 \langle a^{\dag } a \rangle + 1 \right]
\label{ricvarnum}\:.
\end{eqnarray}
This means that the variance of the optimized kernel 
$\overline{\Delta K^2[a^\dag a]}$ becomes much closer to the intrinsic 
intensity fluctuations $\langle\widehat{\Delta n^2}\rangle$ 
than the original noise $\overline{\Delta R^2[a^\dag a]}$.
In order to appreciate such noise reduction we compare the two noise 
ratios
\begin{eqnarray}
\delta n_R = \sqrt{\frac{\overline{\Delta R^2[a^\dag a]}}
{\langle\widehat{\Delta n^2}\rangle}} \qquad
\delta n_K = \sqrt{\frac{\overline{\Delta K^2[a^\dag a]}}
{\langle\widehat{\Delta n^2}\rangle}} \label{ratios}\;.
\end{eqnarray}
For coherent states $|\alpha\rangle$ we obtain
\begin{eqnarray}
\delta n_R = \sqrt{2+\frac{1}{2}\left(|\alpha |^2+\frac{1}{|\alpha |^2}
\right)} \qquad \delta n_K = \sqrt{2+\frac{1}{2|\alpha |^2}}
\label{cratios}\;, 
\end{eqnarray}
that is, from an asymptotically linearly increasing function of $|\alpha |$ 
the ratio becomes a constant $\delta n_K \simeq  \sqrt{2}$. Similar 
expressions are obtained for other kind of state: the noise ratio saturates 
to $\delta n_K \simeq \sqrt{3/2}$ for either squeezed vacuum and cat states. 
\par
\begin{figure}
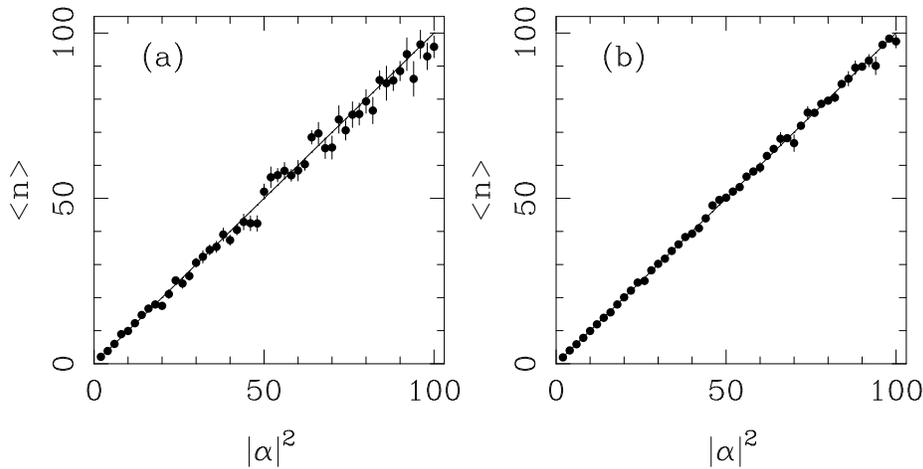

\begin{tabular}{cc}
\psfig{file=adap7a.ps,width=6cm} & \psfig{file=adap7b.ps,width=6cm} \\ 
\end{tabular}
\caption{Tomographic detection of the intensity on coherent states. 
The simulated experiment has been performed with 
15 blocks of 15 data for 15 phases each (for a total number of 
$N= 3375$ measurements). The tomographic result $\langle \hat n \rangle$
is reported versus the theoretical values $|\alpha |^2$, 
(a) without and (b) with optimization. \label{f:avn}} 
\end{figure}

\noindent
In Fig. \ref{f:avn} results from a Monte Carlo simulation of the tomographic 
measurement of intensity on coherent states show the noise reduction obtained
when using the optimized kernel.
\par\noindent 
The noise reduction obtained by adding the single null function 
$F_0(\phi)$  can be easily evaluated also for the generic diagonal moment 
$\langle a^{\dag n} a^n\rangle$, using the formula 
\begin{eqnarray}
e^{i 2\phi} R[a^{\dag n} a^n](x) = \frac{n}{n+1}
\:R[a^{\dag(n+1)} a^{n-1}](x) \: 
\label{form1}\;, 
\end{eqnarray}
which leads to   
\begin{eqnarray}
b_0 = - \overline{ R[a^{\dag n} a^n] e^{i2\phi} }  
= - \frac{n}{n+1} \langle a^{\dag (n+1)} a^{n-1} \rangle 
\label{nnn}\;,
\end{eqnarray}
namely $\Delta^2 [a^{\dag n} a^n]= 2 |b_0|^2$. We just mention 
that optimizing the kernel $R[a^{\dag 2}a^2](x)$ is useful to improve
detection of the second order correlation function $g^{(2)}=
\langle a^{\dag 2}a^2\rangle /\langle a^{\dag}a \rangle^2$.
\subsection{Quadrature}
\begin{figure}
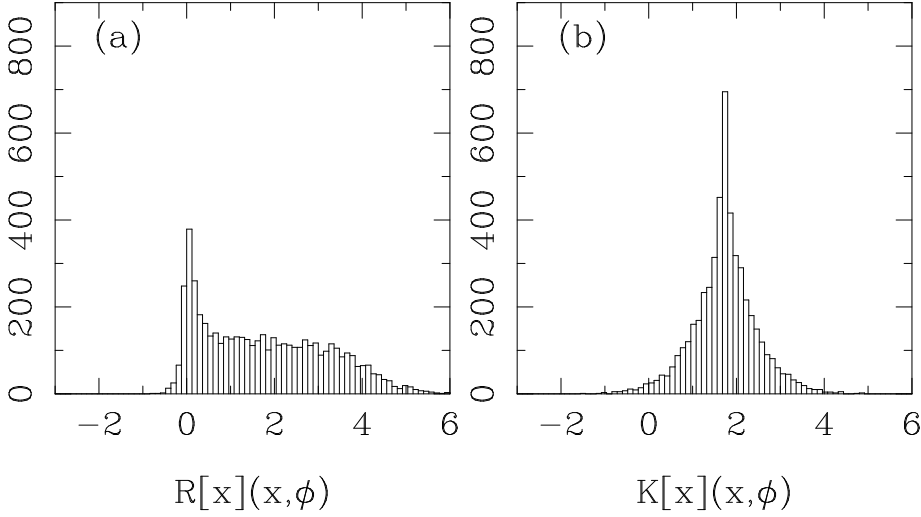

\begin{tabular}{cc}
\psfig{file=adap8a.ps,width=6cm} & \psfig{file=adap8b.ps,width=6cm} \\ 
\end{tabular}
\caption{Histograms of the kernel functions evaluated on the tomographic 
outcomes for a coherent state with $|\alpha|^2 =3$. 
The sample has 50 phases with 100 data each. (a) using the original 
Richter kernel $R[\hat x](x,\phi)$; (b) using the optimized kernel 
$K[\hat x](x,\phi)$ . 
The distribution for the optimized kernel is 
sharper and peaked near the theoretical value 
$\langle\hat x \rangle=\sqrt{3}$. 
\label{f:quad}} 
\end{figure}

The optimization procedure has been tested also on the
kernel $R[\hat x ](x,\phi) = 2 x \cos \phi$, corresponding to the measurement
of the quadrature operator $\hat x =\frac{1}{2}(a+a^\dag )$. 
Similarly to the intensity case, the type-II and type-III null functions
do not play a role in improving precision, whereas type-I functions give
$b_k= - 2^{-k-1}\langle a^{\dag (1+k)}\rangle$ in Eq. (\ref{optlin}).
In this way the optimization procedure can be carried analytically 
also in this case. The results indicate that for 
coherent states it is enough to add the first null function $F_0^I(\phi)$, 
whereas for squeezed vacuum and cat states only  the odd-index functions 
$F_{2s+1}^I(x,\phi)$ contribute to noise reduction. In this case the main 
term is due to $F_1^I(x,\phi)$, whereas higher order functions improve the 
variances only by a few percent. For coherent states the variance reduction 
from $F_0^I(x,\phi )$  is given by
\begin{eqnarray}
\Delta^2 [\hat x ] = \frac{1}{2} \langle a^{\dag }\rangle 
\langle a \rangle = \frac{1}{2} |\alpha |^2 
\label{quadcoh}\;,
\end{eqnarray}
which completely compensates the leading term in the variance of the 
original Richter kernel \cite{add}
\begin{eqnarray}
\overline{\Delta R^2[\hat x ]} =\langle\widehat{\Delta x^2}\rangle 
+ \frac{1}{2} \langle a^{\dag }\:a \rangle + \frac{1}{4}
\label{ricvarquad}\:.
\end{eqnarray}
For squeezed vacuum and cat states the variance reduction due to 
$F_1(x,\phi )$ is 
\begin{eqnarray}
\Delta^2 [\hat x ]= 
\frac{1}{2 \left(1 - |\langle a \rangle|^2 + 2 \langle 
a^\dag a\rangle\right)}\left[|\langle a \rangle|^2 \left(\langle 
a^{\dag 2}\rangle + \langle a^{\dag 2}\rangle + \frac{1}{2}+ \langle 
a^\dag a\rangle\right) + |\langle a^2 \rangle|^2 \right]
\label{long}\:.
\end{eqnarray}
Upon defining the noise ratio $\delta x_K$  in analogy to Eq.(\ref{ratios})
\begin{eqnarray}
\delta x_K = \sqrt{\frac{\overline{\Delta K^2[a^\dag a]}}
{\langle\widehat{\Delta x^2}\rangle}} \label{xratio}\;,
\end{eqnarray} 
from Eqs. (\ref{quadcoh}) and (\ref{long}) we get the constant 
$\delta x_K=\sqrt{2}$ for coherent states, independently on $|\alpha|^2$, 
whereas for squeezed vacuum and cat states the noise ratio saturates to 
$\delta x_K \simeq \sqrt{5/4}$.
In Fig. \ref{f:quad} results from a simulated experiments of tomographic 
measurement of the quadrature on coherent states are shown for $|\alpha
|^2=3$. There the histograms of the original Richter kernel and of  
the optimized kernel are compared. The optimized kernel has a sharper
distribution, which is peaked at the mean value $\langle\hat x \rangle=
\sqrt{3}$. For this reason, it is quite obvious that the optimized kernel
$K[\hat x](x,\phi)$ gives a more precise determination of $\langle\hat x 
\rangle$ than the the original kernel $R[\hat x](x,\phi)$.
\par\noindent 
\subsection{Field amplitude}
The tomographic kernel for the measurement of the complex field amplitude
$a$ is given by $R[a](x,\phi )=2xe^{i\phi}$, and its fluctuations should be 
compared with those from the ideal measurement of $a$, which could be achieved
by ideal eight-port \cite{wal,hai,rip} or six-port \cite{zuc,tri} homodyne 
detection. The optimization procedure depends on the choice for the definition
of statistical error for a complex quantity. If one considers the real or 
the imaginary part separately, the procedure coincides with the 
optimization of the precision in independent measurements of two conjugated 
quadratures. On the other hand, in order to take into account both noises 
jointly, we minimize the quantity 
\begin{eqnarray}
\overline{\Delta_\ast K^2[a]}=\frac{1}{2}\left\{\overline{\left| 
K[a]\right|^2} - \left|\overline{K^2[a]}\right|^2\right\}
\label{newvarc}\;,
\end{eqnarray}
corresponding to the average of noises for real and imaginary 
parts, namely the trace of the noise covariance matrix. 
Now, the equivalence class of kernel functions is written as follows
\begin{eqnarray}
K[a](x,\phi)= R[a](x,\phi) + \sum_{p=0}^{M-1} \mu_p F_p(x,\phi) 
+ \sum_{p=0}^{M-1} \nu_p F_p^*(x,\phi) \label{knullc}\;.
\end{eqnarray}
$\mu_p$ and $\nu_p$ being two independent sets of complex coefficients. 
The optimization procedure is similar to the real case, 
and is reduced to solving the two linear systems 
\begin{eqnarray}
{\bf A} \; {\bf \mu} = {\bf b}  \qquad {\bf A} \; {\bf \nu} = {\bf c} 
\label{optlinc}\;,
\end{eqnarray}
where ${\bf c}$ is given by
$$c_p=- \overline{R[\hat O] F_p}\;.$$
By inverting Eqs. (\ref{optlinc}), one obtains the noise reduction
\begin{eqnarray}
\Delta_\ast^2 [a] =\overline{\Delta_\ast R^2[a]} -
\overline{\Delta_\ast K^2[a]} = \sum_{p,q=0}^{M-1} \left[b_p 
\:\left(A^{-1}\right)_{qp} \:b_q^* + c_p \:\left(A^{-1}\right)_{pq} 
\:c_q^*  \right]\label{redvarc}\;.
\end{eqnarray}
Also in the present case it is sufficient to consider only 
type-I functions. The optimization vector ${\bf b}$ is given by 
$b_k= - 2^{-k}\langle a^{\dag (1+k)}\rangle$. Similarly to the case of 
the quadrature, the optimization procedure shows that for coherent 
states only $F_0^I(\phi)$ is needed, whereas for squeezed vacuum 
and cat states only the odd-index functions $F_{2s+1}^I(x,\phi)$ 
contribute to noise reduction, and the main term comes from $F_1^I(x,\phi)$. 
In this way for coherent states one obtains
\begin{eqnarray}
\Delta_\ast^2 [a ]=  \frac{1}{2} |\alpha|^2
\label{longstarc}\:,
\end{eqnarray}
whereas for squeezed vacuum and cat states one has 
\begin{eqnarray}
\Delta_\ast^2 [a ]= 
\frac{1}{2 \left(1 - |\langle a \rangle|^2 + 2 \langle 
a^\dag a\rangle\right)}\left[|\langle a \rangle|^2 \left(\langle 
a^{\dag 2}\rangle + \langle a^{\dag 2}\rangle + \frac{1}{2}+ \langle 
a^\dag a\rangle\right) + |\langle a^2 \rangle|^2 \right] 
\label{longstar}\:. 
\end{eqnarray}
Eqs. (\ref{longstarc}) and (\ref{longstar}) should be compared with 
the noise-figure of the original Richter kernel
\begin{eqnarray}
\overline{\Delta_\ast^2 R [a]}= 
\frac{1}{2}\left[ 2 \langle a^\dag a \rangle + 1 - |\langle a\rangle|^2
\right]     
\label{ricvarampli}\;,
\end{eqnarray}
and with the intrinsic noise of a generalized measurement of the 
amplitude
\begin{eqnarray}
\langle\widehat{\Delta_\ast a^2}\rangle= 
\frac{1}{2}\left[\langle a^\dag a \rangle + 1 
- |\langle a\rangle |^2\right]     
\label{intri}\;.
\end{eqnarray}
The noise ratios thus equals $\delta a_K=1$ for coherent states, whereas 
saturates to $\delta a_K \simeq \sqrt{3/2}$ for both squeezed vacuum 
and cat states. Remarkably, for coherent 
states the heterodyne noise is reached, namely tomographic detection has
ideal noise.
\section{Effects of systematic errors}\label{s:sys}
\begin{figure}
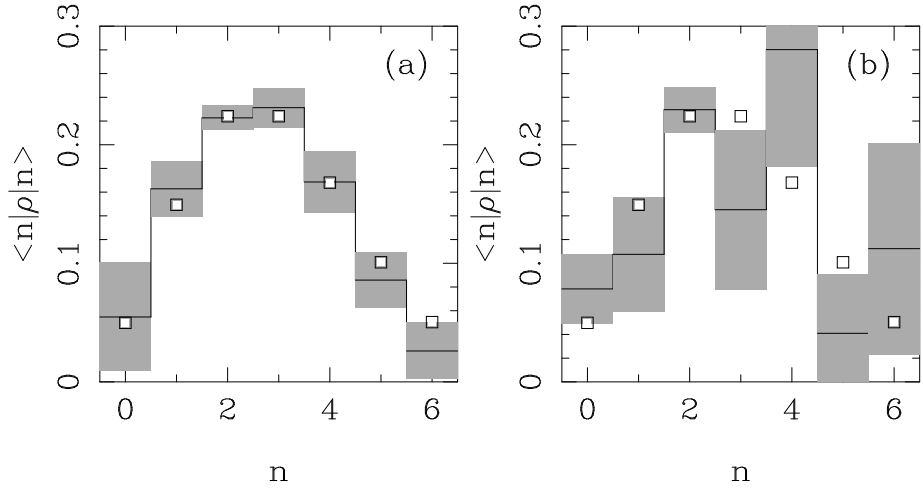

\begin{tabular}{cc}
\psfig{file=adap9a.ps,width=6cm} & \psfig{file=adap9b.ps,width=6cm}
\end{tabular}
\caption{Monte Carlo simulation of adaptive tomography with a bad choice of
the number of added null functions.  The state under examination is a coherent
state with $|\alpha |^2= 3$ and the simulated sample of homodyne data contains
$5$ blocks of $50$ data for $25$ phases each, for a total number of $6250$
measurements (as in Fig. \protect\ref{f:csim}). Here, the optimization has 
been performed by adding $M=32$ null functions. Large fluctuations emerge 
instead of error reduction.  The squares indicate theoretical values.
\label{f:badsim}}
\end{figure}

Throughout this paper the tomographic kernels have been optimized 
by adding low order null functions. Higher order functions oscillate more
rapidly. Since the method involves only the average of these functions
on a small sample of data, fast oscillations in $\phi$ and higher
power of $x$ would introduce more noise, and including too many null 
functions would increase the error instead of reducing it.
In Fig. \ref{f:badsim} an example of such pathology is given. 
\par\noindent
Another point that should be mentioned is that in the tomographic detection 
here considered the phase $\phi$ is a random parameter in $[0,\pi ]$. 
A discrete scanning by equally-spaced phases would introduce systematic 
errors \cite{nic,mun} that would mask the benefits from the 
optimization. 
Actually, for non-random uniform scanning, the null function $F_0(\phi)$ 
has no effects when added to phase independent kernels, whereas 
the other null functions have a much reduced effect, and obviously do not 
eliminate the systematic error due to the finite mesh of the deterministic
scanning. 
\section{Summary and Conclusions}\label{s:out}
In this paper we have presented an adaptive method to optimize tomographic 
kernels, improving the precision of the tomographic measurement.
The method has been analyzed in detail for coherent 
states, Fock states, squeezed vacuum, and "Schr\"{o}dinger-cat" states.
With the exception of Fock, states the method generally provides 
a sizeable reduction of statistical errors. For coherent states the 
improvement mainly concerns the small-index matrix elements, whereas for 
squeezed vacuum and cat states also far off-diagonal elements are improved.
\par\noindent
The error reduction is much more significant for the measurement 
of intensity, quadrature and field amplitude, where for coherent states, 
squeezed vacuum, and cat states the ratio between tomographic noise and 
uncertainty of the considered observable saturates for increasing energy. 
In this case, we can definitely assert that quantum tomography is a
quasi-ideal measurement, as it adds only a small amount of noise as 
compared to ideal detection.
\section*{Acknowledgments}
We would thank Dirk -G. Welsch, Mohamed Dakna and Nicoletta Sterpi 
for useful discussions. M. G. A. Paris has been partly supported by 
the ``Francesco Somaini'' foundation. This work is part of the INFM 
contract PRA-1997-CAT. 

\begin{table}
\caption{Representation table for indices of type-$III$ null functions.
\label{t:III}} 
\begin{tabular}{|c|ccccccccc|}
l  &0&1&2&3&4&5&6&7&.. \\
\hline
k+n &0&1&1&2&2&2&3&3&..\\ 
\hline
k  &0&1&0&2&1&0&3&2&..\\ 
n  &0&0&1&0&1&2&0&1&..\\ 
\end{tabular}\end{table}
\end{document}